\input harvmac
 \noblackbox

 \newcount\figno
 \figno=0
 \def\fig#1#2#3{
 \par\begingroup\parindent=0pt\leftskip=1cm\rightskip=1cm\parindent=0pt
 \baselineskip=11pt
 \global\advance\figno by 1
 \midinsert
 \epsfxsize=#3
 \centerline{\epsfbox{#2}}
 \vskip 12pt
 {\bf Fig.\ \the\figno: } #1\par
 \endinsert\endgroup\par
 }
 \def\figlabel#1{\xdef#1{\the\figno}}
 \def\encadremath#1{\vbox{\hrule\hbox{\vrule\kern8pt\vbox{\kern8pt
 \hbox{$\displaystyle #1$}\kern8pt}
 \kern8pt\vrule}\hrule}}
 %
 %


 \font\cmss=cmss10
 \font\cmsss=cmss10 at 7pt
 \def\rlx{\relax\leavevmode}
 \def\inbar{\vrule height1.5ex width.4pt depth0pt}
 \def\IC{\relax\,\hbox{$\inbar\kern-.3em{\rm C}$}}
 \def\IN{\relax{\rm I\kern-.18em N}}
 \def\IP{\relax{\rm I\kern-.18em P}}
 \def\ZZ{\rlx\leavevmode\ifmmode\mathchoice{\hbox{\cmss Z\kern-.4em Z}}
  {\hbox{\cmss Z\kern-.4em Z}}{\lower.9pt\hbox{\cmsss Z\kern-.36em Z}}
  {\lower1.2pt\hbox{\cmsss Z\kern-.36em Z}}\else{\cmss Z\kern-.4em
  Z}\fi}
 \def\IZ{\relax\ifmmode\mathchoice
 {\hbox{\cmss Z\kern-.4em Z}}{\hbox{\cmss Z\kern-.4em Z}}
 {\lower.9pt\hbox{\cmsss Z\kern-.4em Z}}
 {\lower1.2pt\hbox{\cmsss Z\kern-.4em Z}}\else{\cmss Z\kern-.4em
 Z}\fi}
 \def\IZ{\relax\ifmmode\mathchoice
 {\hbox{\cmss Z\kern-.4em Z}}{\hbox{\cmss Z\kern-.4em Z}}
 {\lower.9pt\hbox{\cmsss Z\kern-.4em Z}}
 {\lower1.2pt\hbox{\cmsss Z\kern-.4em Z}}\else{\cmss Z\kern-.4em Z}\fi}

 \def\narrowplus{\kern -.04truein + \kern -.03truein}
 \def\narrowminus{- \kern -.04truein}
 \def\narrowminussub{\kern -.02truein - \kern -.01truein}

 \def\frac#1#2{{#1\over #2}}

 \def\mp{m_{{\rm P}}}

 \def\IZ{\relax\ifmmode\mathchoice
 {\hbox{\cmss Z\kern-.4em Z}}{\hbox{\cmss Z\kern-.4em Z}}
 {\lower.9pt\hbox{\cmsss Z\kern-.4em Z}}
 {\lower1.2pt\hbox{\cmsss Z\kern-.4em Z}}\else{\cmss Z\kern-.4em Z}\fi}
 \def\IB{\relax{\rm I\kern-.18em B}}
 \def\IC{{\relax\hbox{$\inbar\kern-.3em{\rm C}$}}}
 \def\ID{\relax{\rm I\kern-.18em D}}
 \def\IE{\relax{\rm I\kern-.18em E}}
 \def\IF{\relax{\rm I\kern-.18em F}}
 \def\IG{\relax\hbox{$\inbar\kern-.3em{\rm G}$}}
 \def\IGa{\relax\hbox{${\rm I}\kern-.18em\Gamma$}}
 \def\IH{\relax{\rm I\kern-.18em H}}
 \def\II{\relax{\rm I\kern-.18em I}}
 \def\IK{\relax{\rm I\kern-.18em K}}
 \def\IP{\relax{\rm I\kern-.18em P}}

 \font\cmss=cmss10 \font\cmsss=cmss10 at 7pt
 \def\IR{\relax{\rm I\kern-.18em R}}

 %

 %
 %
 \def\eqnn#1{\xdef #1{(\secsym\the\meqno)}\writedef{#1\leftbracket#1}%
 \global\advance\meqno by1\wrlabeL#1}
 \def\eqna#1{\xdef #1##1{\hbox{$(\secsym\the\meqno##1)$}}
 \writedef{#1\numbersign1\leftbracket#1{\numbersign1}}%
 \global\advance\meqno by1\wrlabeL{#1$\{\}$}}
 \def\eqn#1#2{\xdef #1{(\secsym\the\meqno)}\writedef{#1\leftbracket#1}%
 \global\advance\meqno by1$$#2\eqno#1\eqlabeL#1$$}

\lref\hv{K.~Hori and C.~Vafa,
``Mirror symmetry,''
arXiv:hep-th/0002222.
}
\lref\wphases{
E.~Witten,
``Phases of N = 2 theories in two dimensions,''
Nucl.\ Phys.\ B {\bf 403}, 159 (1993)
arXiv:hep-th/9301042.
}
\lref\dv{
R.~Dijkgraaf and C.~Vafa,
``Matrix models, topological strings, and supersymmetric gauge theories,''
arXiv:hep-th/0206255;
``On geometry and matrix models,''
arXiv:hep-th/0207106;
``A perturbative window into non-perturbative physics,''
arXiv:hep-th/0208048.}
\lref\mp{D.~R.~Morrison and M.~Ronen Plesser,
``Summing the instantons: Quantum cohomology and mirror symmetry in toric varieties,''
Nucl.\ Phys.\ B {\bf 440}, 279 (1995)
arXiv:hep-th/9412236.
}
\lref\other{
L.~Chekhov and A.~Mironov,
``Matrix models vs. Seiberg-Witten/Whitham theories,''
arXiv:hep-th/0209085. 
}
\lref\doret{
N.~Dorey, T.~J.~Hollowood, S.~P.~Kumar and A.~Sinkovics,
``Exact superpotentials from matrix models,''
arXiv:hep-th/0209089; 
``Massive Vacua of N=1* Theory and S-duality from Matrix Models,''
arXiv:hep-th/0209099. 
}

\Title
 {\vbox{
 \baselineskip12pt
 \hbox{hep-th/0209138}\hbox{HUTP-02/A046}\hbox{}
}}
 {\vbox{
 \centerline{Perturbative Derivation of Mirror Symmetry}
 }}

 \centerline{ Mina ${\rm Aganagic}$ and
Cumrun ${\rm Vafa}$}
 \bigskip\centerline{ Jefferson Physical Laboratory}
 \centerline{Harvard University}
\centerline{Cambridge, MA 02138}
 \smallskip
 \vskip .3in \centerline{\bf Abstract}
 {We provide a purely perturbative (one loop) derivation of mirror symmetry for
supersymmetric sigma models in two dimensions.}
 \smallskip \Date{September, 2002}

\newsec{Introduction}

In \hv\ a proof of mirror symmetry was provided in terms of
equivalence of linear sigma models and certain Landau-Ginsburg
theories. In the proof non-perturbative physics, i.e. generation
of superpotential by vortices, played a crucial role. In this
brief note we provide a $perturbative$ derivation of mirror symmetry,
which can be viewed as a purely
perturbative reinterpretation of the
derivation of \hv .

The main motivation to revisit the proof of mirror symmetry is based
on the recent results in
 \dv , where
it was shown that instanton effects in
massive ${\cal N}=1$ supersymmetric theories in four dimensions can be
evaluated in perturbation theory (see also the followup work
\refs{\other , \doret}).
Moreover it was shown that this allows
one to derive non-perturbative S-dualities for 4d supersymmetric
field theories from a
perturbative
perspective.  This leads naturally to the question
of whether one can also derive mirror symmetry
for two dimensional supersymmetric sigma models, which
involves non-perturbative physics, using
only perturbative techniques.  We will show that this is indeed the case.

\newsec{Perturbative Derivation of Mirror Symmetry}

Consider, for example, a $d=2$ supersymmetric sigma model for a
non-compact toric manifold which can be realized as an
${\cal N}=(2,2)$ linear sigma model with one abelian vector
multiplet and $n$ chiral fields charged under it \wphases.
The arguments below easily generalize to the case with
more than one $U(1)$ gauge group.
The local
geometry is specified by the charges of chiral fields $Q_i$,
$i=1,\ldots n$ and the FI term $t$. Provided that
charges sum up to zero $\sum_i Q_i=0$ the theory is expected
to flow to a conformal theory corresponding to a Calabi-Yau
target geometry. In such a case,
the Higgs-branch theory is a non-compact Calabi-Yau
manifold $X$ with $h_{1,1}(X) = 1$, and Kahler class proportional
to $t$.  We will consider, as in \hv , mirror symmetry
in a more generalized sense which includes non-Ricci flat target
geometries as well.

It is well known that the quantum cohomology ring
for supersymmetric sigma models on Kahler manifolds
with $c_1>0$ can be understood from purely perturbative
perspective.  For example for $\ {\IC}P^n$, which can
be realized as a $U(1)$ linear sigma model with $n+1$ charged
fields, integrating out the charged fields leads to a one loop generation
of the superpotential \ref\dadda{A. D'Adda, A.C. Davis, P. Di Vecchia and P. Salomonson,
``An Effective Action for the Supersymmetric $CP^{N-1}$ Sigma Model'',
Nucl.\ Phys.\ B {\bf 222}, 45 (1983)}
$$W=\Sigma ( {\rm log} {\Sigma^{n+1}-(n+1)}) +t \Sigma$$
where $\Sigma$ is the twisted chiral superfield containing the
gauge field.  This also leads to the quantum cohomology
ring
$$dW=0 \rightarrow \Sigma^{n+1}=e^{-t},$$
as is well known, where $\Sigma$ plays the role
of the Kahler class.   Note that  as long
as $\Sigma\not=0$ the charged fields pick up a mass
$M_i\sim Q_i \Sigma$ and can be integrated out.
The fact that extremization
of the superpotential leads to a non-vanishing value
of $\Sigma$ makes integrating out
the charged fields a self-consistent scheme.
 The generalization of the above superpotential
 and quantum cohomology to other $c_1>0$
Kahler manifolds has been studied in \mp .

For the conformal case, where $\sum_i Q_i=0$ integrating
out the charged fields
would not be completely justified, as $\Sigma=0$
is not dynamically ruled out and thus this does not give a
complete self-consistent
description of the theory.  However,
as discussed in \hv\ it is also natural to consider twisted
mass deformations of this theory given by
weakly gauging the global symmetries while
freezing the corresponding vector multiplet scalars to fixed non-zero
values (for earlier work on these deformations see
\ref\twm{L. Alvarez-Gaume and D.Z. Freedman, ``Potentials For The
Supersymmetric Nonlinear Sigma Models'', Commun. Math. Phys. {\bf 91}, 87 (1983);
S.J Gates, ``Superspace Formulation of New Nonlinear Sigma Models'',
Nucl.\ Phys.\ B {\bf 238}, 349 (1984); A. Hanany and K. Hori, ``Branes and N=2
Theories in Two Dimensions'', Nucl.\ Phys.\ B {\bf 513}, 119 (1998),
arXiv:hep-th/9707192.}).  Now,
integrating out the massive charged fields is justified and
the exact twisted superpotential is generated at one loop
for the dynamical vector multiplet $\Sigma$. For simplicity of
notation define \eqn\sigma{\Sigma_i = Q_i \Sigma + m_i,} where
$m_i$ is the twisted mass of the charged fields. The superpotential is, up
to a constant, \dadda
,
\eqn\sup{W(\Sigma) = \sum_i \Sigma_i (log\Sigma_i - 1) +  t_i \Sigma_i,}
where $t_i=t/nQ_i$. The path
integral (which we present only schematically)
\eqn\patha{Z(m,t) = \int {\cal D}\Sigma\;\; e^{W(\Sigma)}}
can be rewritten in terms of integral over $\Sigma_i$ with a delta
function that freezes the non-dynamical part: this should
constrain $\Sigma_i$ to satisfy $n-1$ linear relations that imply
\sigma
\eqn\con{\sum_i  R_i^A \Sigma_i = m^A.}
That is, $R$'s are orthogonal to $Q$'s
$$\sum_i R_i^A Q_i= 0,$$
for $A=1,...,n-1$ and the $n-1$ physical mass terms are given by $m^A$.
The constraints \con\ can be imposed by introducing new twisted chiral fields
$Y_A$
in the theory, playing the role of Lagrange multipliers for
the constraints.  These fields are such that if we integrate
them out we get back the original theory with the same superpotential.
In particular we consider the action
$$Z(m,t) = \int \prod_{i,A} {\cal D} \Sigma_i{\cal D}
Y_A \;\;e^{\sum_i \Sigma_i(log \Sigma_i - 1) + t_i \Sigma_i}\;\;
                                   e^{\sum_{A,i} Y_A (R_i^A \Sigma_i- m_A)},$$
which as far as the F-terms are concerned imposes the above
constraints.
We are not interested in the potential deformations
of the D-terms, and only keep track of the F-terms.
In particular there would also be D-terms involving the $Y_A$ fields which
do not affect the F-terms and we have suppressed them
in the above expression.
It is important to notice that $Y_A$ are $\;\;{\IC}^*$ valued, i.e.
$$Y_A \sim Y_A + 2 i \pi.$$
This follows from the observation made in
\lref\nesh{A. Losev, N. Nekrasov,
and S. Shatashvili, ``The Freckled Instantons'',
arXiv:hep-th/9908204; ''Freckled Instantons in Two and Four
Dimensions'', Class. Quant. Grav. {\bf 17},
1181 (2000), arXiv:hep-th/9911099.}
\refs{\wphases,\nesh} about the nature of the chiral field $\Sigma$.
Recall that the top component of $\Sigma$ contains the field strength
of the $U(1)$ vector field,
$\Sigma =\ldots + \theta^{+}{\bar \theta}^{-} F$.
If we consider the theory on
a compact Riemann surface, then $F$ is quantized,
and consequently Lagrange multiplier enforcing \con\ on $F$ must be periodic
with period $2 \pi$. From the coupling
$\int d\theta^{+} d \bar{\theta}^{-} Y \Sigma = Im(Y) F$
in the superpotential we conclude that it is the imaginary part of $Y$ that
is periodic, and integrating over $Y$'s above, we recover the original
formulation of the theory, \patha.

The path integral over $\Sigma_i$ can be done and localizes on $
\partial_{\Sigma_i}W=0$
             $$log(\Sigma_i) + t_i + \sum_A R_i^A Y_A =0.$$
This gives
$$Z(m,t) = \int \prod_A DY_A \;\;e^{ -\sum_i e^{-Y_i} -  \sum_A Y_A m_A},$$
where we defined $$Y_i =  t_i + \sum_A R_i^A Y_A.$$

This Landau-Ginsburg theory is the known result for the mirror of a
massive deformation of the local A-model theory \hv.

This generalizes to the derivation of mirrors of other local and compact
models. As explained in \hv\ mirror symmetry in the
case of compact manifolds is closely related to the non-compact case
(see also \mp ).
As there are no new ingredients, we refer the reader to \hv\ for detailed
discussion of mirror symmetry in the compact case.

In fact in \hv , by T-dualizing
the charge fields to $Y_A$, it was noticed that in massive cases,
the theory formulated in terms of $(Y_A,\Sigma)$ fields, gives
rise to
the expected superpotential in terms of $\Sigma$ fields
by integrating out the $Y_A$ fields.   In the derivation
we have presented here the $Y_A$  play the role of Lagrange multiplier
fields.  However it is not too difficult to show that the insertion
of the square of the charge fields $|\Phi_i|^2$ is equivalent
to the insertion of $Re Y_i$ which is compatible with the derivation
of \hv .
In this sense, the derivation above is
not new, but it provides a novel perturbative perspective.

\centerline{\bf Acknowledgements}
This work was supported in part by NSF grants PHY-9802709 and DMS-0074329.
\listrefs
 \end